\documentclass[aps,prd,twocolumn,superscriptaddress,preprintnumbers,floatfix,longbibliography]{revtex4-1}

\usepackage{graphicx}
\usepackage{amsmath}
\usepackage[caption=false]{subfig}
\usepackage{siunitx}
\usepackage{placeins}
\usepackage{color}
\usepackage{standalone}
\usepackage{dcolumn}
\usepackage{bm}
\usepackage{microtype}
\usepackage{etoolbox}
\usepackage{amssymb}
\usepackage{mathrsfs}
\usepackage{accents}
\usepackage{gensymb}
\usepackage{float}
\usepackage[normalem]{ulem}
\usepackage[dvipsnames]{xcolor}
\usepackage[colorlinks,urlcolor=NavyBlue,citecolor=NavyBlue,linkcolor=NavyBlue,pdfusetitle]{hyperref}
\usepackage{subfig}

\newtoggle{commentsoff}

\ifdefined\nocomments
\toggletrue{commentsoff}
\fi

\iftoggle{commentsoff}{
	\newcommand{\todo}[1]{}
	\newcommand{\ls}[1]{}
	\newcommand{\dhj}[1]{}
}{
\newcommand*{\todo}[1]{{\color{red} [{\bf TODO}: #1]}}
\newcommand{\ls}[1]{{\color{red}~\textsf{[{\bf LS}: #1]}}}
\newcommand{\dhj}[1]{{\color{blue}~\textsf{[{\bf DHJ}: #1]}}}
}

\newcommand*{\change}[1]{{{#1}}}
\begin{document}

\title{Search for continuous gravitational waves from Fomalhaut b in the second \\
	Advanced LIGO observing run with a hidden Markov model}

\author{Dana Jones}
\affiliation{\mbox{College of Arts and Sciences, University of Pennsylvania, Philadelphia, Pennsylvania 19104, USA}}

\author{Ling Sun}
\affiliation{LIGO Laboratory, California Institute of Technology, Pasadena, California 91125, USA}
\affiliation{OzGrav-ANU, Centre for Gravitational Astrophysics, College of Science, The Australian National University, ACT 2601, Australia}

\date{\today}

\begin{abstract}
Results are presented from a semicoherent search for continuous gravitational waves from a nearby neutron star candidate, Fomalhaut b, using data collected in the second observing run of Advanced LIGO. The search is based on a hidden Markov model scheme, capable of tracking signal frequency evolution from the star's secular spin down and stochastic timing noise simultaneously. The scheme is combined with a frequency domain matched filter ($\mathcal{F}$-statistic), calculated coherently over five-day time stretches. The frequency band 100--1000~Hz is searched. After passing the above-threshold candidates through a hierarchy of vetoes, one candidate slightly above the 1\% false alarm probability threshold remains for further scrutiny. No strong evidence of continuous waves is found. We present \change{the strain upper limits in the full frequency band searched at 90\% confidence level.}
\end{abstract}

\maketitle

\section{Introduction}

Gravitational waves (GWs), perturbations in spacetime that propagate at the speed of light, were first directly observed in 2015 when the Advanced Laser Interferometer Gravitational-Wave Observatory (Advanced LIGO) detected a merging binary black hole system (GW150914) \cite{Abbott2016,LIGO2014}. The Virgo detector joined the observation at the end of the second observing run (O2) in 2017 \cite{Virgo2014}. In the years since the first detection, the sensitivity of these interferometers has been greatly improved, allowing for increasingly frequent detections of compact binary coalescences (CBCs) \cite{Abbott2018,GW190412,GW190425,GW190814}. 
Other types of GW sources that also radiate at frequencies within the observational band of ground-based interferometers remain yet-undetected, e.g., the persistent, well modelled, continuous gravitational waves (CWs) produced by isolated spinning neutron stars. These CWs, if detected, will provide invaluable information regarding the structure of the neutron stars as well as the nuclear equation of state in extreme-pressure situations~\cite{Riles2017}. A great deal of work has been carried out to develop methods and conduct searches for CWs. There are three main types of CW searches: (1) targeted searches for pulsars whose sky positions and ephemerides are well measured electromagnetically (e.g., \cite{Abbott2019-4,Abbott2019-5}), (2) directed searches for neutron stars with known sky positions but unknown rotation frequencies (e.g., \cite{ScoX1cc2017,Abbott2019-2,Abbott2019,O2SNR-Fstat,O2SNR-Viterbi,Ming2019,Fesik2020}), and (3) all-sky searches, surveys done over the whole sky to search for emitting sources (e.g., \cite{Abbott2019-3}). 


In this paper, a directed search, more expensive than a targeted search but less expensive than an all-sky search, is conducted in the Advanced LIGO O2 data \cite{GWOSC} for a nearby neutron start candidate, Fomalhaut b. 
Fomalhaut b was originally hypothesized to be an exoplanet orbiting around the star Fomalhaut A \cite{Currie2012,Kalas1345}. However, certain peculiarities, namely its non-detection in the infrared and its potentially highly eccentric orbit, have led to speculation over whether or not it is in fact an exoplanet. A different hypothesis has been gaining momentum; rather than a companion object, Fomalhaut b may in fact be a background neutron star to Fomalhaut A \cite{Neuhauser2015}. There has also been recent evidence that Fomalhaut b may be a directly imaged catastrophic collision between two large planetesimals in an extrasolar planetary system \cite{Fomalhautb_2020}. 
Although the remaining uncertainty regarding the source's identity may make it a less promising candidate for a CW search, Fomalhaut b remains of interest due to its extremely close proximity. 
It is currently hypothesised to be just 11~pc away, which would make it, assuming the neutron star hypothesis to be true, the closest known neutron star to our solar system \cite{Neuhauser2015}. A CW search directed at Fomalhaut b was conducted in the first observing run of Advanced LIGO, but no evidence of gravitational radiation was found~\cite{Abbott2019}. 

Since the expected strain amplitudes of CWs are orders of magnitude smaller than those produced by CBCs, vast computational resources are required to integrate observational data coherently over a long period of time (e.g., $\sim1$~yr), searching for the signal frequency and the time derivatives~\cite{Riles2017}. In addition, intrinsic, stochastic frequency wandering, or ``timing noise," associated with the secular spin down of a star could degrade the sensitivity of a fully coherent search \cite{Hobbs2010}. Thus, although fully coherent searches are still of interest in certain systems with negligible timing noise, we conduct a computationally efficient semicoherent search based on a hidden Markov model (HMM) scheme, equipped to track the evolving signal frequency due to secular spin down and stochastic timing noise~\cite{Sun2018-2}. 
The tracking scheme has its origins in engineering, and has recently been used in many CW searches (e.g.,~\cite{ScoX1ViterbiO1,Sun2019,Abbott2019-2,O2SNR-Viterbi,Sun2020-CygX1,Sun20191f2f}).
In this search, we assume that the signal frequency evolution is dominated by the star's secular spin down, allowing for minor stochastic timing noise, as the star is an isolated source (cf. the signal evolution is expected to be dominated by timing noise if the source is in an accreting binary system).

The search presented in this paper is conducted using the Advanced LIGO O2 data collected from January 3 to August 25, 2017 in the frequency band 100--1000~Hz, divided into 1-Hz sub-bands to parallelize the computation. The total observing duration is split into five-day coherent segments in order to balance search sensitivity and computing cost in the presence of timing noise. The short Fourier transforms (SFTs) of the data are passed through a frequency domain matched filter (the $\mathcal{F}$-statistic). These coherent five-day segments are combined incoherently using a HMM tracking scheme. The search output is then passed through a hierarchy of veto validations. No strong evidence of continuous waves is found.

The organization of the paper is as follows. Section~\ref{sec:methods} outlines the methods used in the search. Section~\ref{sec:setup} details the search setup, including source parameters, search configuration, threshold, and sensitivity estimates. In Section~\ref{sec:results}, we explain the results from the search, including the five vetoes applied to the over-threshold candidates, and discuss future work that could be used to further follow up on the results. \change{We also present the upper limits obtained on the signal strain and source properties.} Finally, the conclusion is given in Section~\ref{sec:conclusion}.

\section{Methods}
\label{sec:methods}

This search is composed of two main procedures: (1) coherently summing up the signal power over consecutive five-day time stretches using the $\mathcal{F}$-statistic, and (2) a HMM tracking to find the most probable signal evolution path over the total observing run. The signal model is briefly reviewed in Section~\ref{sec:signal_model}. The two procedures are described in Sections~\ref{sec:f-stat} and \ref{sec:HMM}, respectively. The detection statistic adopted in this search, the Viterbi score, is defined in Section~\ref{sec:score}.

\subsection{Signal model}
\label{sec:signal_model}

The phase of the signal as observed in the detector can be described as~\cite{Jaranowski1998}
\begin{equation}
    \label{eqn:phase}
    \Phi(t) = 2\pi \sum_{k=0}^{s} \frac{f_0^{(k)}t^{k+1}}{(k+1)!}+ \frac{2\pi}{c}\hat{n}\cdot\vec{r}(t)\sum_{k=0}^{s}\frac{f_0^{(k)}t^k}{k!},
\end{equation}
where $f_0$ is the signal frequency at reference time $t=0$, the superscript $(k)$ denotes the $k$th time derivative of the signal frequency, $\hat{n}$ is the unit vector directed outward from the solar system barycenter (SSB) to the neutron star, and $\vec{r}(t)$ is the position vector of the detector relative to the SSB. Then, the signal can be expressed as
\begin{equation}
    \label{eqn:sig_form}
    h(t)=\mathcal{A}^\mu h_\mu(t),
\end{equation}
where $\mathcal{A}^\mu$, depending on the characteristic gravitational-wave strain amplitude $h_0$, source orientation and signal initial phase, represents the amplitudes associated with the four linearly independent components~\cite{Jaranowski1998}
\begin{eqnarray}
\label{eqn:h1}
h_1(t) &=& a(t) \cos \Phi(t), \\
h_2(t) &=& b(t) \cos \Phi(t), \\
h_3(t) &=& a(t) \sin \Phi(t),  \\
\label{eqn:h4} h_4(t) &=& b(t) \sin \Phi(t).
\end{eqnarray}
In (\ref{eqn:h1})--(\ref{eqn:h4}), $a(t)$ and $b(t)$ are the antenna-pattern functions given by Eqs.~(12) and (13) in Ref.~\cite{Jaranowski1998}, and $\Phi(t)$ is the signal phase in \eqref{eqn:phase}.

\subsection{$\mathcal{F}$-statistic}
\label{sec:f-stat}

The $\mathcal{F}$-statistic is a matched filter used to estimate the likelihood that a signal described above is present in the frequency domain. The time-domain data $x(t)$ collected by the detector can be written as
\begin{equation}
    x(t)=\mathcal{A}^\mu h_\mu(t) + n(t)
\end{equation}
where $n(t)$ is stationary, additive noise \cite{Jaranowski1998}. First, a scalar product as a sum over single-detector inner products is defined as
\begin{eqnarray}
    (x|y) &=& \mathop{\sum} \limits_{X} (x^X|y^X) \\
    &=& \mathop{\sum} \limits_{X} 4\Re \int_{0}^{\infty}df \frac{\tilde{x}^X(f)\tilde{y}^{X*}(f)}{S_h^X(f)},
\end{eqnarray}
where $X$ indexes the detector, $S_h^X(f)$ is the single-sided power spectral density (PSD) of detector $X$, the tilde denotes a Fourier transform, and $\Re$ is the real part of a complex number \cite{Prix2007}. Then, the $\mathcal{F}$-statistic can be written in the form
\begin{equation}
    \mathcal{F} = \frac{1}{2} x_\mu \mathcal{M}^{\mu \nu} x_\nu,
\end{equation}
where $x_\mu = (x|h_\mu)$, and $\mathcal{M}^{\mu \nu}$ represents the matrix inverse of $\mathcal{M}_{\mu \nu}=(h_\mu|h_\nu)$ \cite{Cutler2005}. If the noise is Gaussian and the single-sided PSD is the same in all detectors, the probability of having a signal in the data solely depends the signal-to-noise ratio, given by~\cite{Jaranowski1998}
\begin{equation}
    \label{eqn:snr2}
    \rho_0^2=\frac{K h_0^2T_{\rm coh}}{S_h(f)},
\end{equation}
where $K$ is a constant that depends on the sky location, orientation of the source, and number of detectors, and $T_{\rm coh}$ is the length of data combined coherently.

\subsection{Hidden Markov model}
\label{sec:HMM}

This semicoherent search, based on the HMM scheme, employs the Viterbi algorithm to identify the most likely frequency evolution path of the signal. (See Ref.~\cite{Viterbi1967} for an explanation of the classic Viterbi algorithm.) It is computationally efficient and robust in the presence of timing noise~\cite{Sun2018-2}.

A Markov chain is a stochastic process that transitions from one discrete state to another at discrete times. A hidden Markov model is comprised of two variables: the unobservable, hidden state variable \mbox{$q(t) \in \{q_1, \cdots, q_{N_Q}\}$} and the observable, measurement state variable \mbox{$o(t) \in \{o_1, \cdots, o_{N_O}\}$}, where $N_Q$ and $N_O$ are the total number of hidden and measurement states, respectively. For any time $t_{n+1}$, the hidden state is solely dependent on the state at time $t_n$ and has a transition probability of
\begin{equation}
	\label{eqn:transition_prob}
	A_{q_j q_i} = \Pr [q(t_{n+1})=q_j|q(t_n)=q_i].
\end{equation}
The hidden state $q_i$, present in the observable state $o_j$, has an emission probability defined as
\begin{equation}
	L_{o_j q_i} = \Pr [o(t_n)=o_j|q(t_n)=q_i],
\end{equation}
at time $t_n$.
The prior is written as
\begin{equation}
	\Pi_{q_i} = \Pr [q(t_1)=q_i],
\end{equation}
where $t_1$ is the reference time of the first time step.
The probability that an observed sequence \mbox{$O = {o(t_1),...,o(t_{N_T})}$}, where $N_T$ is the total number of time steps, is the result of a hidden state path $Q = {q(t_1),...,q(t_{N_T})}$ via a Markov chain can be described by
\begin{equation}
	\begin{split}
		\Pr(Q|O) \propto & L_{o(t_{N_T})q(t_{N_T})} A_{q(t_{N_T})q(t_{N_T-1})} \cdots L_{o(t_2)q(t_2)} \\ 
		& \times A_{q(t_2)q(t_1)} \Pi_{q(t_1)}.
	\end{split}
\end{equation}
The most probable path, calculated by maximizing $\Pr(Q|O)$ is  \cite{Sun2018-2}
\begin{equation}
	Q^*(O)= \arg\max \Pr(Q|O),
\end{equation}
where $\arg\max(\cdots)$ returns the argument that maximizes~$(\cdots)$.

In this search, the one-dimensional state variable $q(t)$ is defined as $f_0(t)$. The discrete hidden states are mapped one-to-one to the frequency bins that make up the output of $\mathcal{F}(f)$ calculated over the span of length $T_{\rm coh}$ (see Section~\ref{sec:f-stat}). Each frequency bin size is then $\Delta f = 1/(2T_{\rm coh})$. We choose $T_{\rm coh}$ to satisfy 
\begin{equation}
    \left|\int_t^{t+T_{\rm coh}}dt' \dot{f_0}(t')\right| < \Delta f
    \label{eqn:Tcoh}
\end{equation}
for $0<t<T_{\rm obs}- T_{\rm coh}$, where $ T_{\rm obs}$ is the total observation time. (See Ref.~\cite{Sun2018-2} for more details).

Assuming that the frequency evolution caused by timing noise is much slower than that due to the star's secular spin down, and $|\dot{f}_0 (t)|$ lies in the range between zero and the maximum estimated spin-down rate $|\dot{f}_{0}|_{\rm max}$, the signal frequency evolution can be approximated by a negatively biased random walk with $|\dot{f}_0 (t)| \in [0, |\dot{f}_{0}|_{\rm max}]$. By substituting $|\dot{f}_{0}|_{\rm max}$ into \eqref{eqn:Tcoh}, we can simplify \eqref{eqn:transition_prob} to become
\begin{equation}
    A_{q_{i-1} q_i} = A_{q_i q_i} = \frac{1}{2},
\end{equation}
with all other $A_{q_j q_i}$ entries vanishing. Then, using the definition of $\mathcal{F}$-statistic, the emission probability is defined as
\begin{eqnarray}
    L_{o(t) q_i} &=& \Pr [o(t)|{f}_i \leq f_0(t) \leq {f}_i+\Delta f] \\ 
    &\propto& \exp[\mathcal{F}({f}_i)],
\end{eqnarray}
from $t$ to $t + T_{\rm coh}$, where $f_i$ represents the central frequency in the $i$th bin. A uniform prior of $\Pi_{q_i} = N_Q^{-1}$ is selected because there is no independent knowledge of $f_0$~\cite{Sun2018-2}.

\begin{figure*}
	\centering
	\includegraphics[scale=.32]{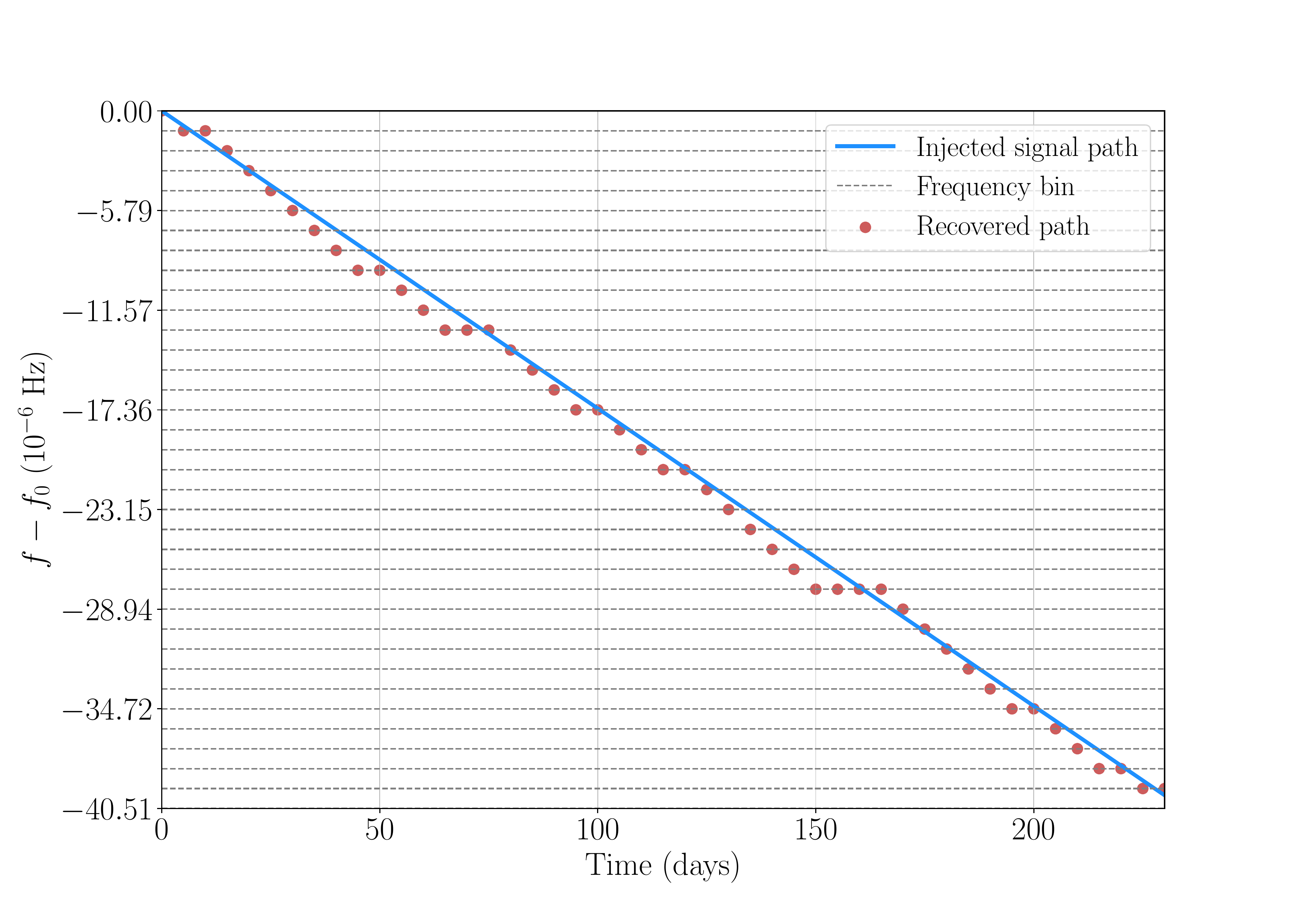}
	\caption{HMM tracking of a synthetic signal with $h_0 = 1.57 \times 10^{-26}$ and $\cos \iota = 1$, injected into Gaussian noise (ASD $S_h^{1/2} = 4 \times 10^{-24}$~Hz$^{-1/2}$) at a starting frequency of 155.3~Hz. The blue curve represents the injected signal (with $\dot{f_0}=-2\times 10^{-12}$~Hz\,s$^{-1}$ and $\ddot{f_0}=2\times 10^{-24}$~Hz\,s$^{-2}$), and the red dots represent the optimal Viterbi path recovered by the HMM tracking. The dashed lines indicate the size of the discrete frequency bins.}
	\label{fig:HMM}
\end{figure*}

The algorithm outputs the most likely frequency evolution path $Q^*(O)$ over the course of $T_{\rm obs}$. This is called the Viterbi path and consists of a frequency estimated at each discrete time step. 
Figure~\ref{fig:HMM} shows an example of tracking a circularly polarized synthetic signal with $h_0 = 1.57 \times 10^{-26}$ and $\cos \iota = 1$, where $\iota$ is the source inclination angle, starting at $f_0=155.3$~Hz (a frequency chosen randomly), and injected into Gaussian noise with amplitude spectral density (ASD) \mbox{$S_h^{1/2} = 4 \times 10^{-24}$~Hz$^{-1/2}$}. The total observing time tracked is 234~days, with $T_{\rm coh}=5$~d. The first and second time derivatives of the injected signal frequency are $\dot{f_0}=-2\times 10^{-12}$~Hz\,s$^{-1}$ and $\ddot{f_0}=2\times 10^{-24}$~Hz\,s$^{-2}$. The optimal signal evolution path reconstructed by the Viterbi algorithm (red dots) is plotted over the injected signal path (blue curve).
The reconstructed path shows a stair-step pattern because the HMM method uses discrete frequency bins (shown as dashed lines). The injected signal, by contrast, evolves continuously. The recovered path matches the injected signal well in that it never strays more than two bins from the true frequency.
The root-mean-square error between the two paths is $7.8 \times 10^{-7}$~Hz, smaller than the discrete bin size, $1.16 \times 10^{-6}$~Hz. 

\subsection{Viterbi score}
\label{sec:score}
We follow the existing literature and define Viterbi score to evaluate the significance of any candidate output from the search~\cite{ScoX1ViterbiO1,Sun2018-2}.
In each sub-band searched (with width 1~Hz in this paper), the Viterbi score $S$ is defined such that the log likelihood of the optimal Viterbi path is equal to the mean log likelihood of all paths ending in different bins of the sub-band plus $S$ standard deviations at final step $N_T$. This is shown as follows:
\begin{equation}
\label{eqn:viterbi_score}
S = \frac{\ln \delta_{q^*}{(t_{N_T})} -\mu_{\ln \delta}(t_{N_T})}{\sigma_{\ln \delta}(t_{N_T})}
\end{equation}
where
\begin{equation}
\mu_{\ln \delta}(t_{N_T}) = N_Q^{-1} \sum_{i=1}^{N_Q} \ln \delta_{q_i}(t_{N_T})
\end{equation}
and
\begin{equation}
\sigma_{\ln \delta}(t_{N_T})^2 = N_Q^{-1} \sum_{i=1}^{N_Q} [\ln \delta_{q_i}(t_{N_T}) - \mu_{\ln \delta}(t_{N_T}) ]^2.
\end{equation}
Here, $\delta_{q_i}(t_{N_T})$ is the maximum probability of the path that ends in state $q_i$ ($1\leq i \leq N_Q$) at step $N_T$, and $\delta_{q^*}{(t_{N_T})}$ is the likelihood of the optimal Viterbi path (i.e., the probability $\Pr[Q^*(O)|O]$). The higher the score $S$, the more likely that a signal is present in the sub-band.

\section{Search setup} 
\label{sec:setup}

In this section, we first discuss the source parameters and the parameter space covered in the search in Section~\ref{sec:parameters}. Sections~\ref{sec:threshold} and \ref{sec:sensitivity} describe the detection threshold and estimated search sensitivity, respectively. 

\subsection{Source parameters and search configuration}
\label{sec:parameters}

Fomalhaut b is located at right ascension 22~h 57~m 39.1~s and declination $29^\circ 37' 20.0''$ (J2000). Ref.~\cite{Abbott2019} considered both the most optimistic and pessimistic estimates of Fomalhaut b's distance (0.011--0.02~kpc) and age (316--3000~kyr) in the literature.

Here, we mainly rely on the estimated, age-based spin-down rate in order to choose an optimal $T_{\rm coh}$ in this search. 
The age-based $\dot{f_0}$ is estimated using \cite{Sun2018-2, Abbott2019}
\begin{equation}
-\frac{f_0}{(n_{\rm min}-1) t_{\rm age}} \leq \dot{f_0} \leq -\frac{f_0}{(n_{\rm max}-1) t_{\rm age}},
\label{eqn:fdot}
\end{equation}
where $t_{\rm age}$ is the age of the source, and $n_{\rm min}$ and $n_{\rm max}$, respectively, represent the minimum and maximum values of the breaking index $n=f_0 \ddot{f_0}/\dot{f_0}^2$. 
Given an estimated $\dot{f_0}$, a $T_{\rm coh}$ value may then be chosen based on the following equation:
\begin{equation}
T_{\rm coh} \leq (2|\dot{f_0}|)^{-1/2},
\label{eqn:Tcoh1}
\end{equation}
such that \eqref{eqn:Tcoh} is satisfied. Since the search sensitivity improves as $T_{\rm coh}$ increases for a given $T_{\rm obs}$ \cite{Sun2018-2}, we usually set $T_{\rm coh} = (2|\dot{f_0}|)^{-1/2}$.

The frequency range searched in this analysis is 100--1000~Hz, where the Advanced LIGO detectors are most sensitive.
Table~\ref{tab:rang} lists the $|\dot{f_0}|$ ranges calculated using \eqref{eqn:fdot}, and the corresponding $T_{\rm coh}$ ranges calculated using \eqref{eqn:Tcoh1}, for $f_0 \in [100, 1000]$~Hz. The ranges are calculated for each $n \in \{2,5,7\}$ and $t_{\rm age} \in \{316, 3000\}$~kyr. 
Since the uncertainty of the estimated $t_{\rm age}$ is large, we select an intermediate $T_{\rm coh}=5$~d, corresponding to a $\dot{f_0}$ range of $\dot{f_0} \in [-2.68 \times 10^{-12}, 0] $~Hz\,s$^{-1}$.
With this choice of $T_{\rm coh}$, the search can cover most of the interesting parameter space, if the torque is dominated by gravitational radiation reaction ($n=5$) or $r$-mode oscillations ($n=7$), e.g., almost the full 100--1000~Hz band if $t_{\rm age} \sim 10^3$~kyr, or the most sensitive hundred-hertz band if $t_{\rm age} \sim 10^2$~kyr. 
Signals with $\dot{f_0}$ out of the covered range during part of the observing run could still be partially tracked by the HMM. The desired sensitivity of the search (Section~\ref{sec:sensitivity}), however, cannot be achieved.
Note that although the minimum $T_{\rm coh} = 0.82$~d in Table~\ref{tab:rang} could cover all scenarios listed above, it is not the optimal choice since the sensitivity degrades as $T_{\rm coh}$ decreases ($\propto T_{\rm coh}^{-1/4}$)~\cite{Sun2018-2}. 

\begin{table}
	\centering
	\setlength{\tabcolsep}{5pt}
	\renewcommand\arraystretch{1.5}
	\begin{tabular}{llll}
		\hline
	    n & $t_{\rm age}$ (kyr) & $|\dot{f}_{0}|$ $(\rm{Hz\,s^{-1}})$ & $T_{\rm coh}$ (days) \\
		\hline
		2 & 316 & [$1.00 \times 10^{-11}$, $1.00\times 10^{-10}$] & [0.82, 2.58] \\
		2 & 3000 & [$1.06 \times 10^{-12}$, $1.06 \times 10^{-11}$] & [2.52, 7.96] \\
		
		5 & 316 & [$2.51 \times 10^{-12}$, $2.51\times 10^{-11}$] & [1.63, 5.17] \\
		5 & 3000 & [$2.64 \times 10^{-13}$, $2.64 \times 10^{-12}$] & [5.03, 15.92] \\
		
		7 & 316 & [$1.67 \times 10^{-12}$, $1.67\times 10^{-11}$] & [2.00, 6.33] \\
		7 & 3000 & [$1.76 \times 10^{-13}$, $1.76 \times 10^{-12}$] & [6.17, 19.50] \\
		\hline
	\end{tabular}
	\caption{Estimated ranges of $|\dot{f_0}|$ and $T_{\rm coh}$. The frequency range considered is 100--1000~Hz.}
	\label{tab:rang}
\end{table}

The total observation period searched is the duration of Advanced LIGO's O2 run, excluding the first few weeks when the data quality was not optimal and was followed by a brief end-of-the-year break; that is, GPS time 1167545066--1187733592 (January 3 to August 25, 2017), a total duration of about 234 days. Given the chosen $T_{\rm coh}=5$~d, we have the total number of steps $N_T = 47$.

\subsection{Threshold}
\label{sec:threshold}
The Viterbi score threshold $S_{\rm th}$, corresponding to a desired false alarm probability $\alpha_{\rm f}$ in each sub-band (\mbox{$\alpha_{\rm f}=1\%$} in this search), is determined through Gaussian noise simulations. 
In each of the five sample 1-Hz sub-bands, starting from 100~Hz, 300~Hz, 500~Hz, 700~Hz, and 900~Hz, 200 realizations of pure Gaussian noise are generated and searched with the same configuration as used in the real search (i.e., $T_{\rm coh}=5$~d and $T_{\rm obs}=234$~d). The resulting 1000 scores are sorted and then the score at the 99th percentile is determined as the threshold, $S_{\rm th}=6.95$. Ten out of the 1000 scores obtained from pure noise are above $S_{\rm th}$, corresponding to $\alpha_{\rm f}=1\%$ per 1-Hz sub-band.

This threshold is verified using O2 interferometric data. In three clean 1-Hz sub-bands, starting from 300~Hz, 600~Hz, and 900~Hz, 200 noise-only realizations are simulated by drawing random sky positions. The total 600 realizations yield a threshold $S_{\rm th}=6.97$ ($\alpha_{\rm f}=1\%$ per 1-Hz sub-band), in good agreement with the threshold obtained in pure Gaussian noise. Hence we set $S_{\rm th}=6.95$ in this search.

\subsection{Sensitivity}
\label{sec:sensitivity}

We evaluate the search sensitivity by quantifying the signal strength $h_0$ required to achieve a 95\% detection efficiency, denoted by $h_0^{95\%}$. It is first obtained by injecting 100 synthetic signals with a fixed $h_0$ value into Gaussian noise (ASD $S_h^{1/2} = 4 \times 10^{-24}$~Hz$^{-1/2}$) in four 1-Hz sub-bands, starting from 155~Hz, 355~Hz, 555~Hz, and 755~Hz. These simulations are marginalized over the orientation of the source, with the sky location fixed at the true position of Fomalhaut b. If more than 95\% of the signals are correctly recovered, this process is repeated with a smaller value of $h_0$, and vice versa. We obtain an average of $h_0^{95\%} = 3.91 \times 10^{-26}$ in these four sub-bands.

This process is repeated in O2 interferometric data in a sample 1-Hz sub-band starting from 715~Hz, randomly chosen from among the relatively clean bands. We obtain $h_0^{95\%} = 1.67 \times 10^{-25}$, an order of magnitude larger than $h_0^{95\%}$ obtained in Gaussian noise. This is because (1) the O2 noise ASD has not reached the design sensitivity of Advanced LIGO, and (2) the duty cycle (i.e., the proportion of time that the data is in analyzable science mode) of O2 is only about 50\%. The sensitivity in the full frequency band is presented in Sec.~\ref{sec:upperlimits} and interpreted as the upper limits on $h_0$.

\section{Results}
\label{sec:results}

In this section, we describe the first-pass candidates obtained from the analysis with $S>S_{\rm th}$, and a series of vetoes validating these candidates in Section~\ref{sec:candidates}. After the veto procedure, only one candidate (with Viterbi score slightly above threshold) remains for further scrutiny. Without strong evidence of CW, we present the strain upper limits and astrophysical interpretation in Section~\ref{sec:upperlimits}.

\subsection{Candidates and vetoes}
\label{sec:candidates}

\begin{figure*}
    \centering
    \includegraphics[scale=.35]{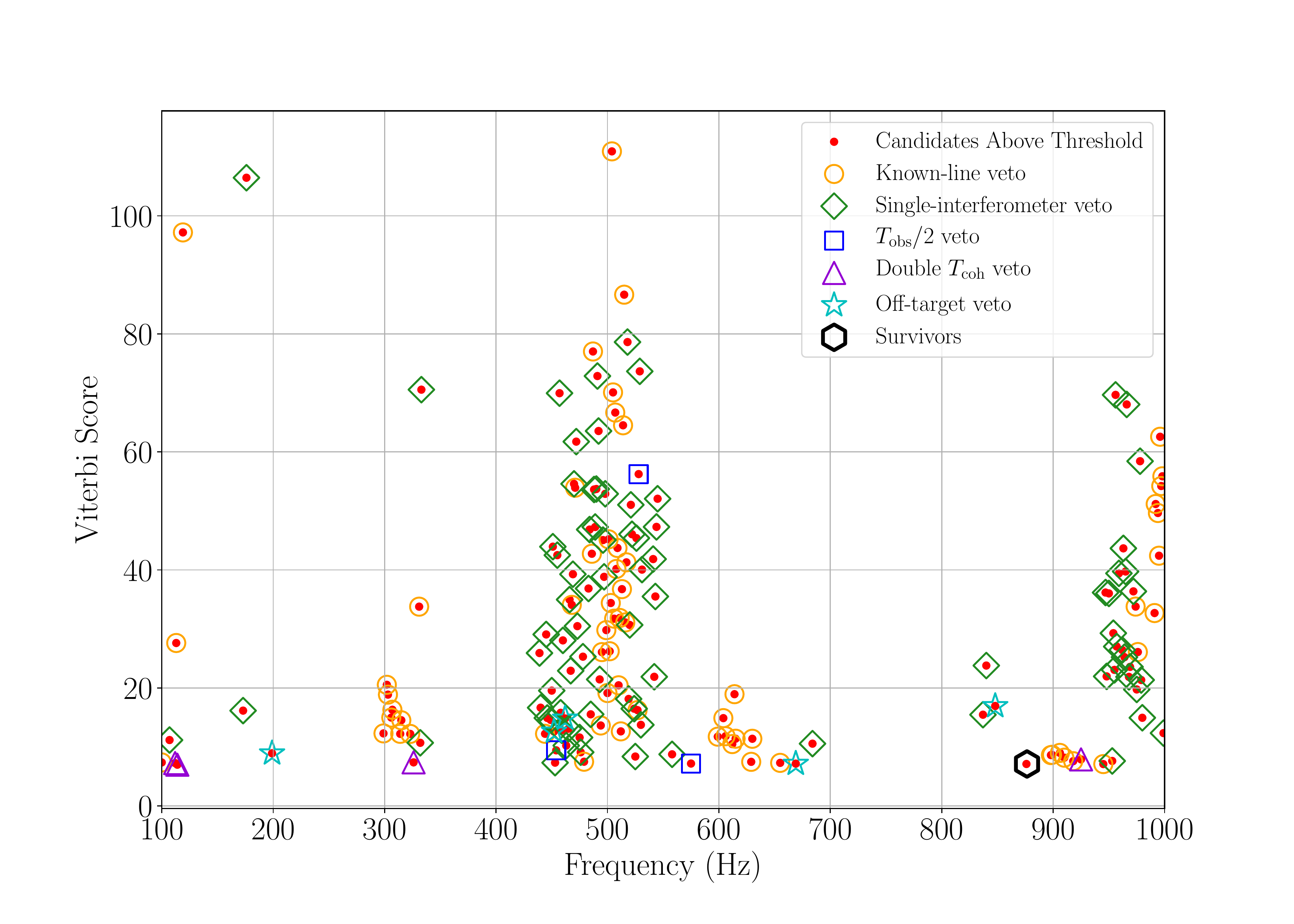}
    \caption{Viterbi score $S$ of the first-pass candidate in each 1-Hz sub-band as a function of frequency (red dots). Candidates marked by orange circles, green diamonds, blue squares, purple triangles, and cyan stars are eliminated in stages 1--5 of the veto procedure (see Section~\ref{sec:candidates} for details of the veto criteria). The only surviving candidate is marked by a black hexagon.}
    \label{fig:vetoes}
\end{figure*}

We find in total 160 first-pass candidates with \mbox{$S>S_{\rm th}$}. 
A large portion of the first-pass candidates are contributed from noise artifacts and the non-Gaussianity in the interferometric data. 
Narrow-band noise lines, e.g., 60~Hz power line harmonics, thermally excited mirror suspension violin modes, and lines from environmental disturbances are all sources of noise artifacts that obscure astrophysical CW signals.
A five-step veto process is conducted to eliminate candidates resulting from noise artifacts \cite{ScoX1ViterbiO1}. See Table~\ref{tab:vetoes} for the number of candidates remaining after each step. See Figure~\ref{fig:vetoes} for all the candidates with their Viterbi scores plotted as a function of frequency, as well as at which step each candidate is vetoed. The detailed veto criteria are described as follows.

\begin{table}[H]
    \centering
    \setlength{\tabcolsep}{10pt}
	\renewcommand\arraystretch{1.5}
    \begin{tabular}{ll}
        \hline
        Processing step & Candidates Remaining \\
        \hline
        First pass & 160 \\
        Know-line veto & 96 \\
        Single-interferometer veto & 13 \\
        $T_{\rm obs}/2$ veto & 10 \\
        Double $T_{\rm coh}$ veto & 6 \\
        Off-target veto & 1 \\
        \hline
    \end{tabular}
    \caption{Number of candidates remaining after each processing step. In total 160 first-pass candidates are found with \mbox{$S>S_{\rm th} = 6.95$}. After the five veto steps, one candidate remains for further scrutiny.}
    \label{tab:vetoes}
\end{table}

First, we take into account the maximum Doppler shift due to the Earth’s motion, $\delta f \approx 10^{-4} f_0$ (here $f_0$ is the starting frequency of the path), and widen the Viterbi path of each candidate by $\delta f$ on both sides of the path. We eliminate 64 candidates whose widened Viterbi paths intersect any known instrumental lines present in either the Hanford or Livingston interferometer~\cite{line-identification,O2lines}.

Second, an additional 83 candidates are vetoed due to the contamination from not well understood artifacts in a single detector. The criteria are as follows. For each candidate, if searching data from a single interferometer yields $S \geq S_{\cup}$, where $S_{\cup}$ is the original score obtained with both interferometers combined, while searching the other interferometer yields $S < S_{\rm th}$, and if the Viterbi path from the interferometer with $S \geq S_{\cup}$ intersects the original path, that candidate is vetoed. We eliminate 49 more candidates here.
The above criterion is a stringent consistency check. After manually checking the remaining candidates, we find that a large portion of them are also caused by artifacts in a single detector, but do not necessarily meet the stringent criterion above. We inspect the candidate scores and paths manually and veto an additional 34 candidates due to contamination from a single detector. They fall into one of the three special cases: (a) We have $S>S_{\cup} \gg S_{\rm th}$ in one detector and $S < S_{\rm th}$ in the other, i.e., the candidate power completely comes from a single detector. The Viterbi path from the interferometer with $S \geq S_{\cup}$ does not intersect the original path (hence not automatically vetoed), but the two paths only differ by $\lesssim 0.02$~Hz.  
(b) We have a larger $S>S_{\cup} \gg S_{\rm th}$ in one detector, but in the other we have $S$ slightly above $S_{\rm th}$ (hence not automatically vetoed). The Viterbi path from the detector with $S>S_{\cup}$ intersects the original path, and the Viterbi path from the detector with $S \sim S_{\rm th}$ does not intersect the original path. In addition, the higher $S$ from one detector is at least a number of 40 larger than $S$ from the other.
(c) We have $S \sim S_{\cup} \gg S_{\rm th}$ in one detector ($S$ slightly below $S_{\cup}$, hence not automatically vetoed) and $S < S_{\rm th}$ in the other. The Viterbi path from the detector with larger $S$ intersects the original path. In this case, the candidate is so loud that the Viterbi score is saturated around $S_{\cup} \gtrsim 50 \gg S_{\rm th}$.

\begin{figure*}[!tbh]
	\centering
	\subfloat[]{\includegraphics[scale=.28]{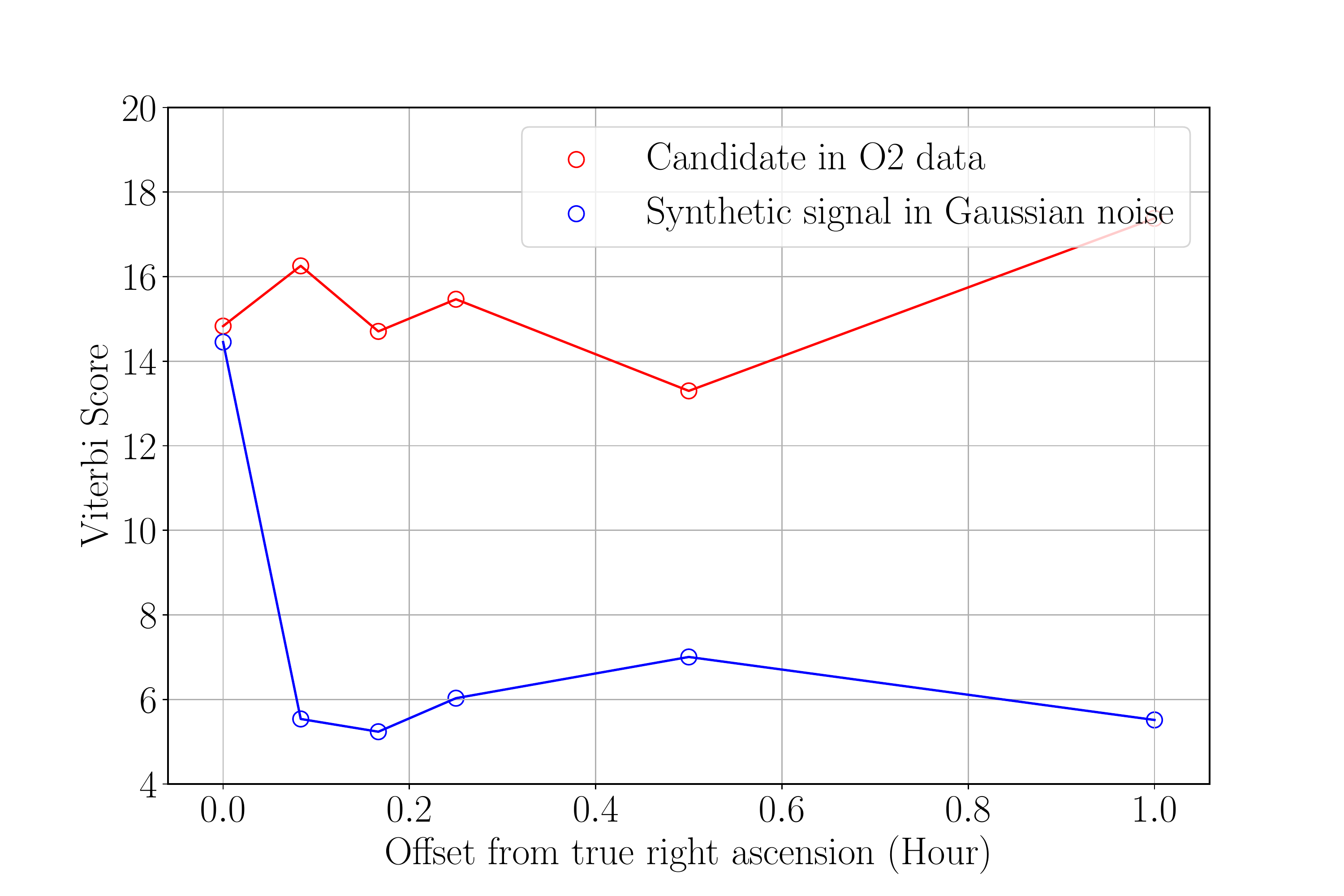}}
	\qquad
	\subfloat[]{\includegraphics[scale=.28]{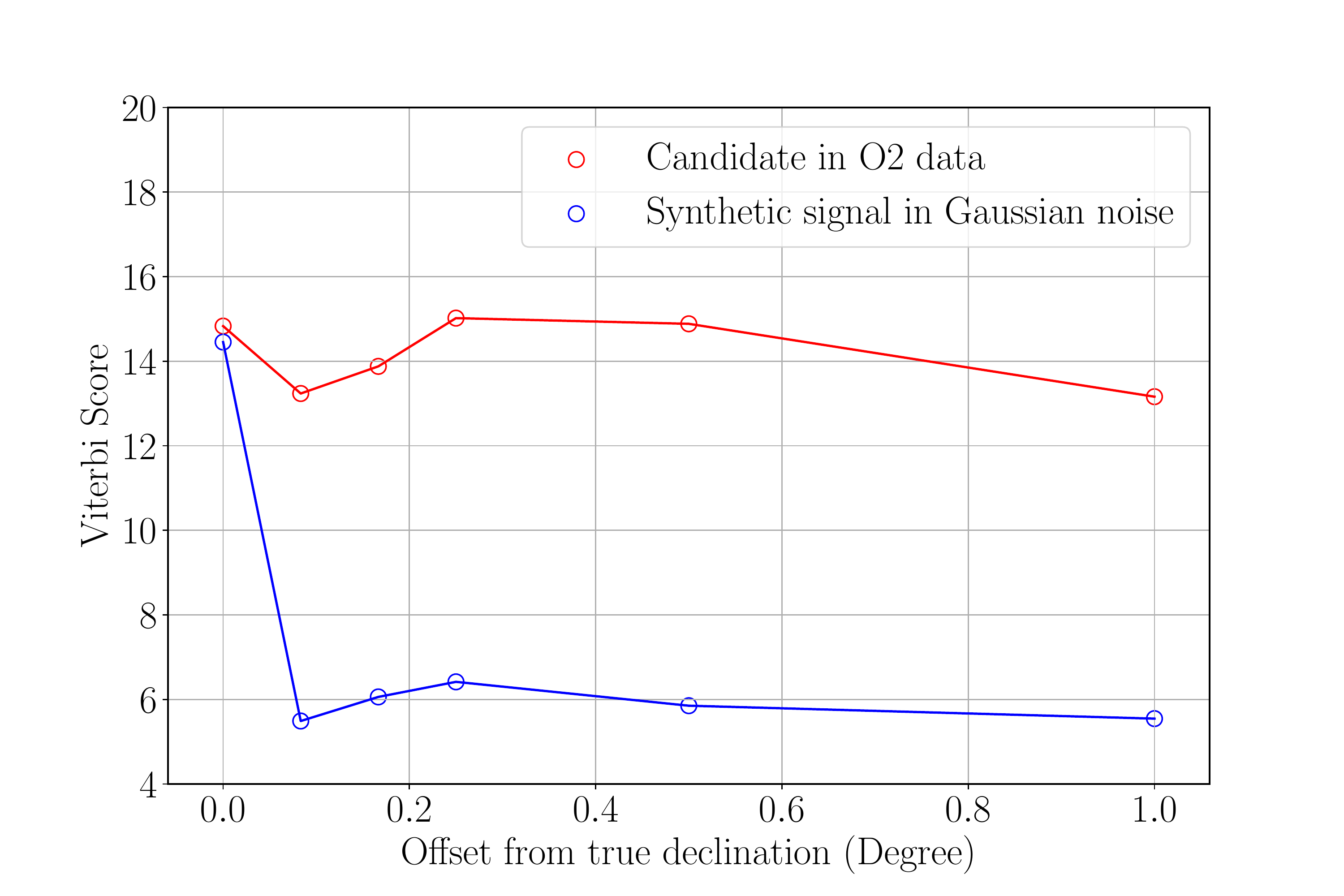}}
	\caption{Viterbi score as a function of the (a) right ascension and (b) declination offsets in the 1-Hz sub-band starting at 462~Hz. The red and blue curves show the results obtained when following up the candidate in the real data and an injection in Gaussian noise, respectively. The results obtained for the candidate in real data show a steady and even increasing score as the offset increases. The simulation shows that the score drops below $S_{\rm th}$ within $\sim 5'$ from the true location (either for right ascension or declination). This candidate can therefore be vetoed.}
	\label{fig:462veto5}
\end{figure*}

\begin{figure*}[!tbh]
	\centering
	\subfloat[]{\includegraphics[scale=.28]{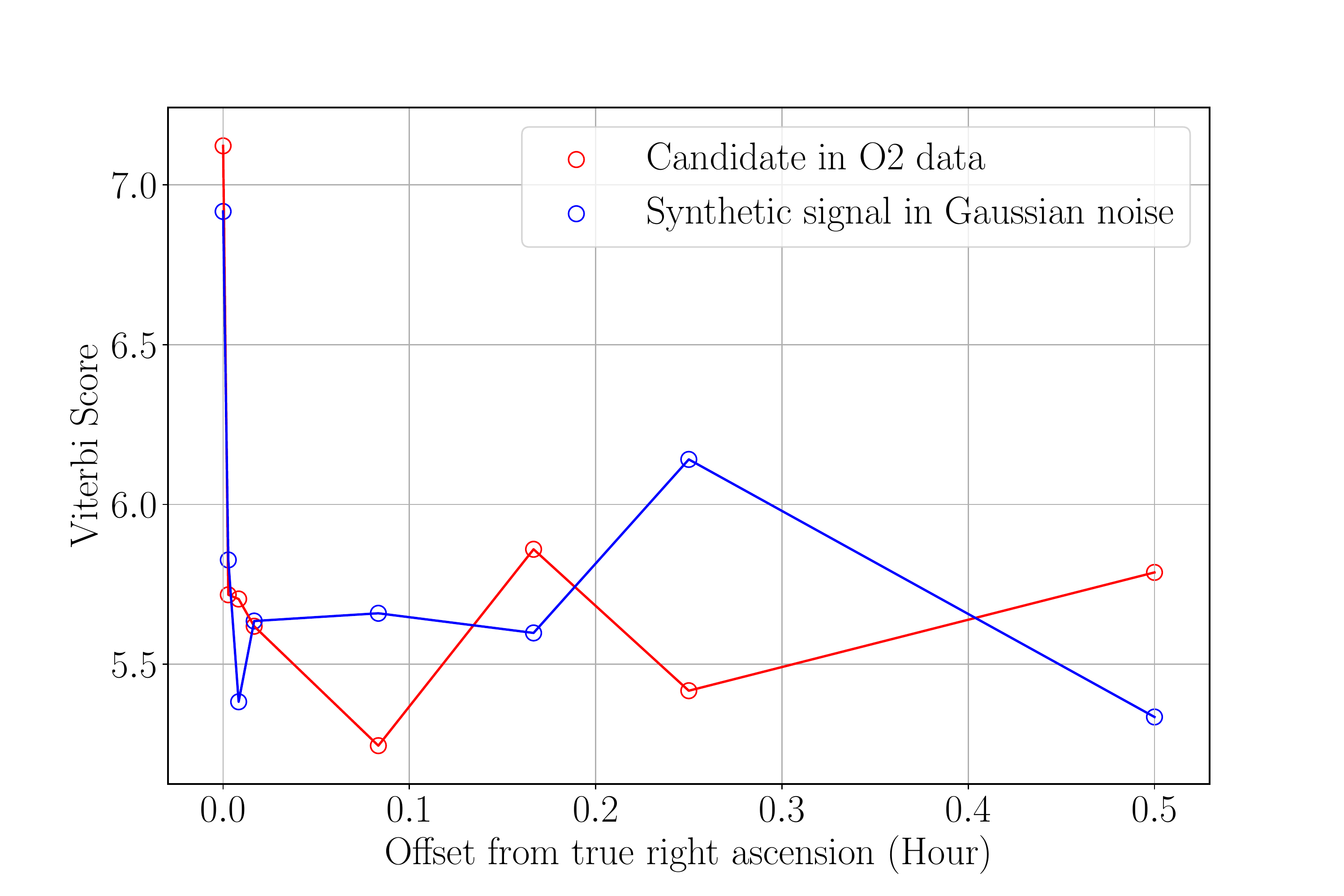}}
	\qquad
	\subfloat[]{\includegraphics[scale=.28]{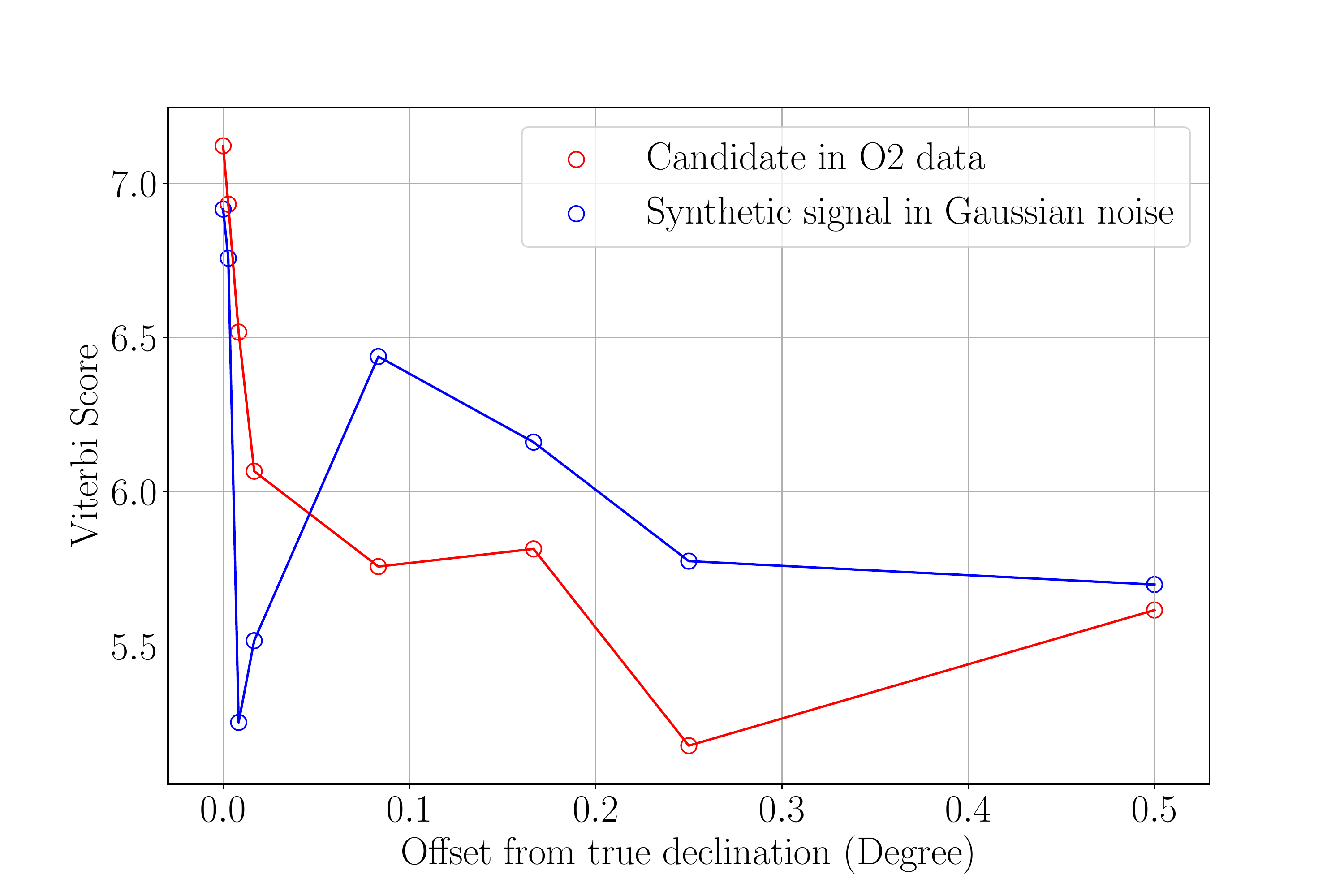}}
	\caption{Viterbi score as a function of the (a) right ascension and (b) declination offsets in the 1-Hz sub-band starting at 876~Hz. The red and blue curves show the results obtained when following up the candidate in the real data and an injection in Gaussian noise, respectively. Both the scores obtained for the candidate in real data and the injection drop below threshold within $\sim 30''$ from the true location. The candidate survives this veto.}
	\label{fig:876veto5}
\end{figure*}

Third, an additional three candidates are vetoed by splitting the observation time into two halves and analyzing each half separately. A candidate is vetoed if the score obtained by searching one half of $T_{\rm obs}$ is higher than or equal to $S_{\cup}$ (corresponding to the full $T_{\rm obs}$), while the score obtained in the other half is lower than $S_{\rm th}$, and if the Viterbi path from the half with the higher score intersects the original path.

Fourth, the remaining ten candidates are further followed up by increasing $T_{\rm coh}$. We increase $T_{\rm coh}$ from five to ten days for a candidate whose estimated mean $\dot{f_0}$ over $T_{\rm obs}$ is small enough such that Equation~\eqref{eqn:Tcoh1} holds true when $T_{\rm coh}=10$~d. Then, for each candidate, the $\mathcal{F}$-statistic and the Viterbi path are recomputed, using both interferometers combined. If the candidate is from astrophysical origin, the resulting Viterbi score with $T_{\rm coh}=10$~d should be higher than $S_{\cup}$, and the output Viterbi paths from searches with $T_{\rm coh}=5$~d and 10~d should match. Hence we veto a candidate if its score corresponding to $T_{\rm coh}=10$~d falls below $S_{\cup}$ and if the Viterbi paths from these two separate searches are at least 0.01~Hz apart. The candidate in the 462~Hz sub-band is excluded from this procedure because its estimated mean $\dot{f_0}$ is so large that \eqref{eqn:Tcoh1} is no longer satisfied if we increase $T_{\rm coh}$ to 10~d. We eliminate four candidates in this step.

Fifth, all but one candidate is eliminated using an ``off-target" veto. For each of the remaining six candidates, we first shift the sky location by increasing the right ascension by offsets ranging from 10~sec to 1~hr from the source's true location, with the declination fixed. 
We then increase the declination by offsets ranging from 10~sec to 1~deg 
from the source's true location, with the right ascension fixed at the true value.
We conduct the search and obtain the Viterbi score for each off-target position. 
A candidate is vetoed if this series of searches targeting sky locations shifted from the source's true location do not continuously yield Viterbi scores lower than the original score. That is, if the score does not drop below the threshold as we move away from the source's true location, we veto the candidate.
The reliability of this veto is verified through synthetic signals injected into Gaussian noise. The same series of searches targeting the true sky location of the injection as well as the off-target locations is conducted. The simulation results demonstrate that the score drops below $S_{\rm th}$ as we move away from the true sky location of the injection with an offset of $\sim 5'$ in either right ascension or declination.  
A sample in the 462--463~Hz sub-band, which corresponds to a vetoed candidate in this step, is shown in Figure~\ref{fig:462veto5}.

The only candidate that survives the off-target veto is in the 876--877~Hz sub-band. The off-target search results in the real data in this sub-band behave exactly like those obtained from the simulations, as shown in Figure~\ref{fig:876veto5}. The Viterbi score immediately drops as we move away from the source's true location. The scores remain below threshold as we continue to increase the offset. Since this candidate behaves like an astrophysical signal, we keep it for further scrutiny.

Finally, the single remaining candidate in the 1-Hz sub-band starting at 876~Hz (path starts at $f_0=876.5034513780022$~Hz) is followed up using a technique in which the Doppler modulation, which accounts for the Doppler shift due to Earth's motion, is turned off when computing $\mathcal{F}$-statistic~\cite{DMoff}, and the Viterbi score is recomputed in the same sub-band. 
A signal of astrophysical origin usually becomes undetectable and a different Viterbi path is returned when turning the Doppler modulation off. 
However, a candidate caused by noise artifacts should yield a higher score and return a Viterbi path close to the original one, i.e., if the original Viterbi path is expanded by a frequency shift due to the Doppler modulation, the expanded path should intersect the new path returned when the Doppler modulation is turned off.
Searching this final candidate with the Doppler modulation turned off yields a decreased Viterbi score 5.91 (cf. 7.12 in the original search). The newly returned Viterbi path differs from the original path by $\sim 0.005$~Hz ($< 10^{-4} f_0$). The two paths are considered overlapped when taking into consideration the frequency shift due to the Earth’s orbit. 
The results for this candidate do not confidently show consistency with either an astrophysical signal or an outlier caused by noise artifacts. 
Therefore, while the candidate cannot be vetoed at this point, the possibility that it comes from astrophysical origin is low. Given the threshold chosen for $\alpha_{\rm f}=1\%$ per sub-band and the fact that the original score of the candidate is only slightly above this threshold ($S-S_{\rm th}=0.17$), the final candidate in sub-band 876--877~Hz could possibly be a false alarm. We provide the full Viterbi path of the final candidate in Appendix~\ref{appendix:candi876} and recommend following it up in future observing runs, including the data already collected in the third observing run.

\subsection{Strain upper limits}
\label{sec:upperlimits}

\begin{figure*}[!tbh]
	\centering
	\includegraphics[scale=.35]{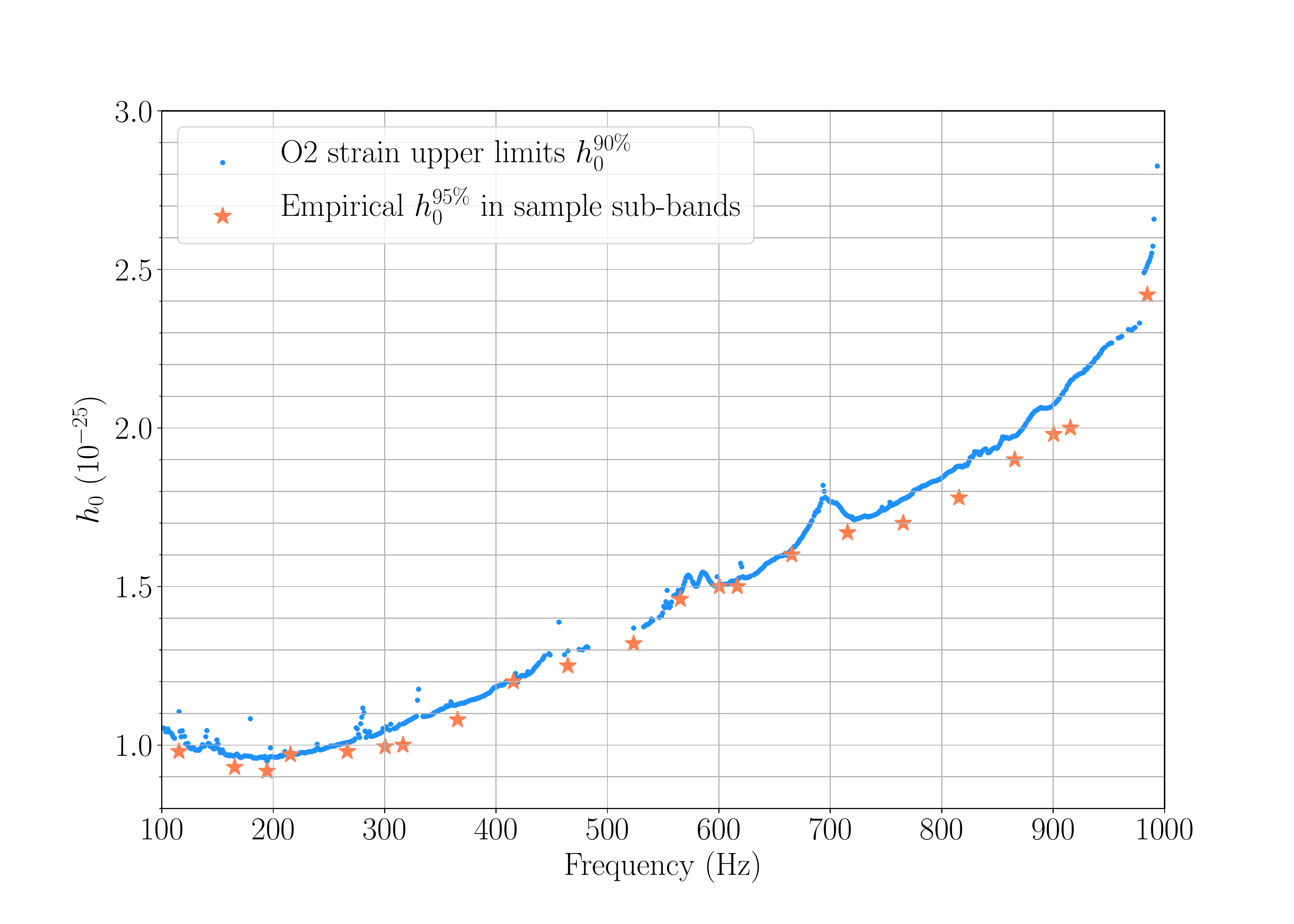}
	\caption{The strain upper limits as a function of frequency. \change{The blue dots represent the 90\% confidence level strain upper limits, $h_0^{90\%}$, in the full band}, and the orange stars represent the 22 sample 1-Hz sub-bands where the $h_0^{95\%}$ values were obtained empirically. The sub-bands without a marker were vetoed.}
	\label{fig:upperlimit}
\end{figure*}

\begin{figure*}[!tbh]
	\centering
	\subfloat[]{\includegraphics[scale=.32]{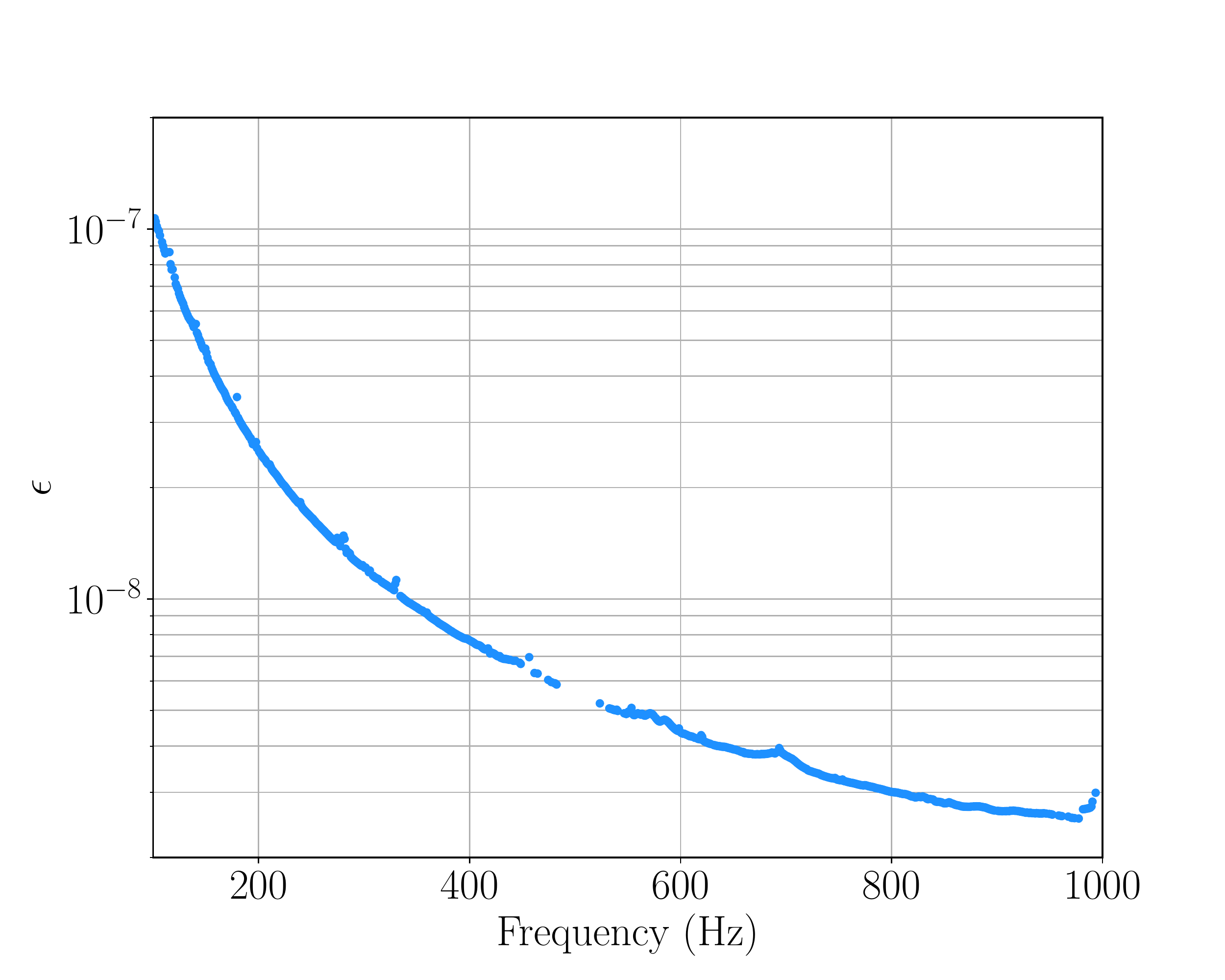}}
	\qquad
	\subfloat[]{\includegraphics[scale=.32]{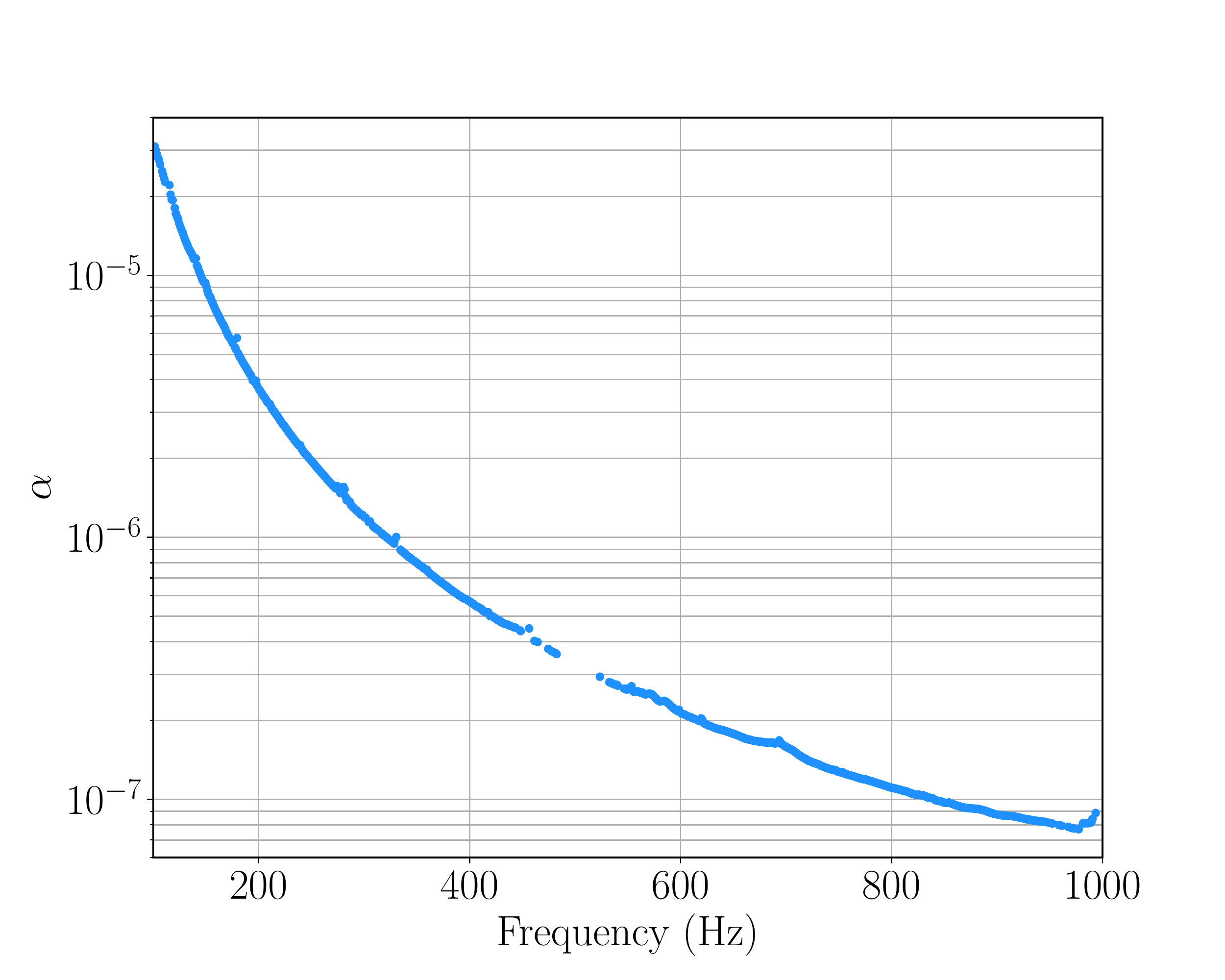}}
	\caption{Upper limits on (a) fiducial ellipticity $\epsilon$ and (b) $r$-mode amplitude $\alpha$, converted from the strain upper limits in Figure~\ref{fig:upperlimit}.}
	\label{fig:eps_alpha}
\end{figure*}

No strong evidence of continuous waves is found in this search. We \change{place} the upper limits on the signal strain at \change{90\%} confidence level, denoted by \change{$h_0^{90\%}$}, as a function of frequency.
Figure~\ref{fig:upperlimit} shows \change{$h_0^{90\%}$} in all 741 sub-bands where no candidate is vetoed (blue dots), as well as 22 sample $h_0^{95\%}$ values obtained from injections (orange stars). The procedure of producing these results is as follows.
First, we derive the upper limits at 95\% confidence level \change{empirically in 22 randomly selected sample sub-bands, marked by the orange stars}. A set of 100 synthetic signals are injected into the O2 data at the sky position of Fomalhaut b in each sub-band with a fixed $h_0$. The other source parameters are randomly drawn from their uniform distributions. The corresponding detection rate is calculated. This process is repeated with different $h_0$ values with step size $1 \times 10^{-27}$ until 95\% of the injections are recovered. 
\change{Next, we calculate the sensitivity depth $\mathcal{D} = \sqrt{S_h(f)}/h_0^{95\%}(f)$ in these 22 sample sub-bands.} 
Here $\sqrt{S_h(f)}$ is the effective ASD calculated from the harmonic mean of the two detectors over all the 30-min SFTs collected from GPS time 1180310418--1187733592 ($\sim3$~months).
\change{In order to save computing cost, we use the minimum depth in the 22 sample sub-bands, $\mathcal{D}_{\rm min}$, as the property of the search pipeline and calculate the strain upper limits in the full band, i.e. $\sqrt{S_h(f)}/\mathcal{D}_{\rm min}$ (the blue dots).}
The sub-bands containing vetoed candidates, where no upper limit can be placed, are excluded.
\change{Finally, we correct for the confidence level of these upper limits calculated in the full band.
We denote the ratio of all the sub-bands satisfying $\mathcal{D} \geq \mathcal{D}_{\rm min}$ to the total 741 sub-bands by $\beta$. 
By construction, we have $\mathcal{D} \geq \mathcal{D}_{\rm min}$ in all 22 randomly chosen sub-bands.
The probability for this to happen in any 22 randomly chosen sub-bands can be approximated by $\beta^{22}$ (when the number of sample sub-bands is much smaller than the total number of sub-bands). By assuming that this probability is at least 95\%, we get $\beta \geq 0.95^{1/22}$. 
In any randomly selected sub-band, there is at least 95\% chance that the signal can be detected if $\mathcal{D} \geq \mathcal{D}_{\rm min}$; the probability of having $\mathcal{D} \geq \mathcal{D}_{\rm min}$ in that sub-band is $\beta$; and the probability for $\beta \geq 0.95^{1/22}$ to be valid is at least 95\%.
Hence the new confidence level in the full band is at least 90\% (i.e., $0.95 \times 0.95^{1/22} \times 0.95$). 
We denote the strain upper limits in the full band by $h_0^{90\%}$.
In any given sub-band, there is a probability lower than 10\% that a signal with $h_0 \geq h_0^{90\%}$ is missed in this search.}

The indirect upper limit on signal strain due to energy conservation, inferred from the age and distance of the source, is given by \cite{Wette2008,Abbott2019}
\begin{equation}
\label{eqn:h0_age}
h_0^{\rm age} = 1.26\times 10^{-24} \left(\frac{3.3\,{\rm kpc}}{d}\right) \left(\frac{300\,{\rm yr}}{t_{\rm age}}\right)^{1/2},
\end{equation}
where $d$ is the distance to the star. Assuming that Fomalhaut b is a neutron star with $t_{\rm age}=3000$~kyr and $d=0.02$~kpc (the pessimistic scenario presented in Ref.~\cite{Abbott2019}), we have $h_0^{\rm age} = 2.08 \times 10^{-24}$. The upper limits obtained in this analysis beat the indirect limit $h_0^{\rm age}$ in the full band searched. 
We compare the strain upper limits obtained from this search to the previous results in Ref.~\cite{Abbott2019}. In the most sensitive band around 200~Hz, $h_0^{95\%} = 1.2 \times 10^{-25}$ is obtained in Advanced LIGO's first observing run, as presented in Ref.~\cite{Abbott2019}. In this search, the strain upper limit in the most sensitive 1-Hz sub-band starting from 194~Hz, \change{$h_0^{90\%} = 9.48 \times 10^{-26}$},
improves on previously published results by a factor of 1.3. 
We also compare our results to the $h_0$ upper limits reported in all-sky searches for CWs using the O2 data. Our most constraining \change{$h_0^{90\%}$} is a factor of 1.8 and 1.4 lower than the best upper limits $h_0^{95\%} \simeq 1.7 \times 10^{-25}$ at around 120~Hz reported in Ref.~\cite{Abbott2019-3}, and $h_0^{90\%} = 1.3 \times 10^{-25}$ at 163~Hz reported in Ref.~\cite{EatHO2}, respectively. At higher frequencies in the range of 500--1000~Hz, our upper limits \mbox{$\sim 1.3 \times 10^{-25}$--$2.7 \times 10^{-25}$} are comparable and slightly better than those \mbox{$\sim 1.5 \times 10^{-25}$--$3 \times 10^{-25}$} reported in an all-sky search targeting low-ellipticity sources~\cite{FalconO2}.

We then convert \change{$h_0^{90\%}$} to the upper limits on the fiducial ellipticity of the neutron star \cite{Wette2008,Abbott2019}
\begin{equation}
\label{eqn:eps}
\epsilon = 9.5\times 10^{-5} \left(\frac{h_0}{10^{-24}}\right) \left(\frac{d}{1\,{\rm kpc}}\right) \left(\frac{100\,{\rm Hz}}{f}\right)^2,
\end{equation}
and the $r$-mode amplitude parameter \cite{Owen2010,Lindblom1998,Abbott2019}
\begin{equation}
\label{eqn:alpha}
\alpha = 0.028 \left(\frac{h_0}{10^{-24}}\right) \left(\frac{d}{1\,{\rm kpc}}\right) \left(\frac{100\,{\rm Hz}}{f}\right)^3.
\end{equation}
Figure~\ref{fig:eps_alpha} shows the upper limits on $\epsilon$ and $\alpha$ as a function of frequency, assuming a distance of $d=0.011$~kpc, and the principal moment of inertia $I_{zz} = 10^{45}$~g\,cm$^2$.

\section{Conclusion}
\label{sec:conclusion}

We report the results from a directed search for CW signals from the neutron star candidate Fomalhaut b in the frequency band 100--1000~Hz, using Advanced LIGO O2 data. An efficient, semicoherent search method, based on a HMM tracking scheme and a matched filter $\mathcal{F}$-statistic, is used to track the signal frequency. After passing the above-threshold candidates through five veto steps, only one candidate survives. The final surviving candidate appears to be consistent with the false alarm probability, and no compelling evidence of continuous waves is found. Upper limits on $h_0$ with \change{90\% confidence are calculated} in the full frequency band. In the most sensitive sub-band, 194--195~Hz, we obtain \change{$h_0^{90\%} = 9.48 \times 10^{-26}$}.

In the future, the search may benefit from new information gathered from electromagnetic observations of the source, as well as the further improvements to the instruments. In particular, the final candidate in the sub-band 876--877~Hz can be followed up in the third observing run or future runs of Advanced LIGO and Virgo. Regardless of the eventual outcome of this search, this methodology represents an important development in the search for CWs and could easily be extended to other isolated neutron star sources.

\section{Acknowledgments}
This research has made use of data, software and/or web tools obtained from the Gravitational Wave Open Science Center (https://www.gw-openscience.org), a service of LIGO Laboratory, the LIGO Scientific Collaboration and the Virgo Collaboration. LIGO is funded by the U.S. National Science Foundation. Virgo is funded by the French Centre National de Recherche Scientifique (CNRS), the Italian Istituto Nazionale della Fisica Nucleare (INFN) and the Dutch Nikhef, with contributions by Polish and Hungarian institutes.
The authors would like to thank the LIGO Laboratory for providing the resources with which to conduct this search, as well as Alan Weinstein and the LIGO SURF program, the National Science Foundation, and the California Institute of Technology for sponsoring the project. 
The authors also would like to thank Meg Millhouse for the review and suggestions.
L. Sun is a member of the LIGO Laboratory. 
LIGO was constructed by the California Institute of Technology and Massachusetts Institute of Technology with funding from the United States National Science Foundation, and operates under cooperative agreement PHY--1764464. Advanced LIGO was built under award PHY--0823459. 
L. Sun also acknowledges the support of the Australian Research Council Centre of Excellence for Gravitational Wave Discovery (OzGrav), project number CE170100004.
The authors would like to thank all of the essential workers who put their health at risk during the COVID--19 pandemic, 
without whom we would not have been able to complete this work.
This paper carries LIGO Document Number LIGO--P2000118. 

\appendix
\section{Viterbi path of the candidate in 876--877~Hz band}
\label{appendix:candi876}

\begin{figure*}
	\centering
	\includegraphics[scale=.32]{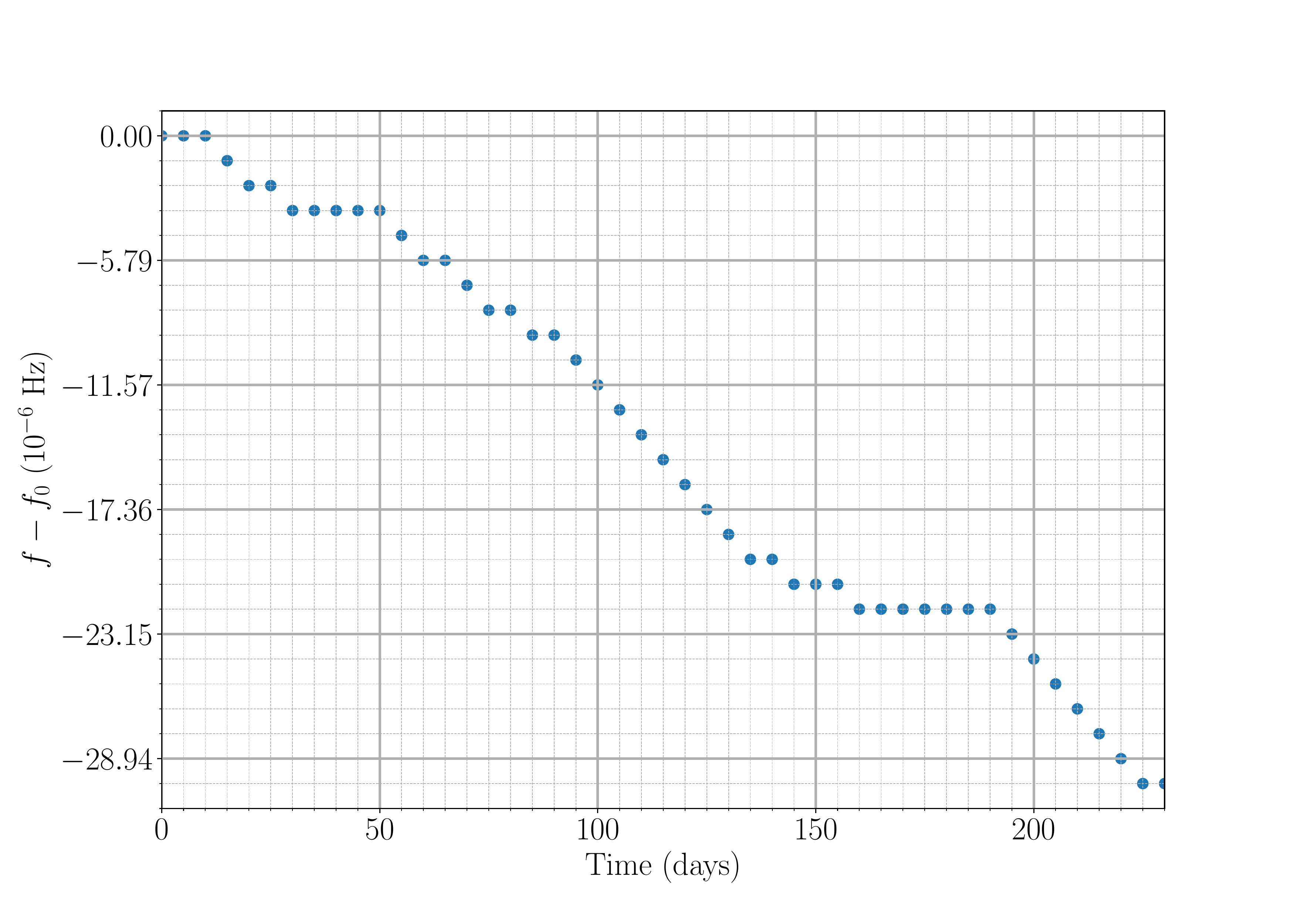}
	\caption{Viterbi path of the final candidate surviving all vetoes in the sub-band 876--877~Hz. The starting frequency equals the frequency at the first step in Table~\ref{tab:876path}, $f_0=876.5034513780022$~Hz.}
	\label{fig:876path}
\end{figure*}

Here we provide the output Viterbi path for the final candidate surviving all vetoes in Section~\ref{sec:candidates}. The frequencies at discrete time steps are listed in Table~\ref{tab:876path} and plotted in Figure~\ref{fig:876path}.

\begin{table}[tbh]
	\centering
	\setlength{\tabcolsep}{10pt}
	\renewcommand\arraystretch{1.2}
	\begin{tabular}{ll}
		\hline
		Step &Frequency (Hz) \\
		\hline
		1&876.5034513780022\\
		2&876.5034513780022\\
		3&876.5034513780022\\
		4&876.5034502205948\\
		5&876.5034490631874\\
		6&876.5034490631874\\
		7&876.5034479057800\\
		8&876.5034479057800\\
		9&876.5034479057800\\
		10&876.5034479057800\\
		11&876.5034479057800\\
		12&876.5034467483727\\
		13&876.5034455909653\\
		14&876.5034455909653\\
		15&876.5034444335579\\
		16&876.5034432761505\\
		17&876.5034432761505\\
		18&876.5034421187431\\
		19&876.5034421187431\\
		20&876.5034409613357\\
		21&876.5034398039284\\
		22&876.5034386465210\\
		23&876.5034374891136\\
		24&876.5034363317062\\
		25&876.5034351742988\\
		26&876.5034340168914\\
		27&876.5034328594841\\
		28&876.5034317020767\\
		29&876.5034317020767\\
		30&876.5034305446693\\
		31&876.5034305446693\\
		32&876.5034305446693\\
		33&876.5034293872619\\
		34&876.5034293872619\\
		35&876.5034293872619\\
		36&876.5034293872619\\
		37&876.5034293872619\\
		38&876.5034293872619\\
		39&876.5034293872619\\
		40&876.5034282298545\\
		41&876.5034270724472\\
		42&876.5034259150398\\
		43&876.5034247576324\\
		44&876.5034236002250\\
		45&876.5034224428176\\
		46&876.5034212854102\\
		47&876.5034212854102\\
		\hline
	\end{tabular}
	\caption{Frequencies at each step on the candidate Viterbi path.}
	\label{tab:876path}
\end{table}

\end{document}